\documentclass[aps,pra,twocolumn, showpacs, superscriptaddress]{revtex4}  

\usepackage{graphicx}  
\usepackage{amsmath}
\usepackage{bm}        
\usepackage{amssymb}   
\usepackage{hyperref}

\newcommand{\be}{\begin{equation}}
\newcommand{\ee}{\end{equation}}
\newcommand{\ben}{\begin{eqnarray}}
\newcommand{\een}{\end{eqnarray}}

\begin{document}

\title{Orbital angular momentum entanglement in turbulence}
\author{A. Hamadou Ibrahim} \affiliation{CSIR National Laser Centre, PO Box 395, Pretoria 0001, South Africa}
\affiliation{University of Kwazulu-Natal, Private Bag X54001, 4000 Durban, South Africa}
\author{Filippus S. Roux} \affiliation{CSIR National Laser Centre, PO Box 395, Pretoria 0001, South Africa}
\author{Melanie McLaren} \affiliation{CSIR National Laser Centre, PO Box 395, Pretoria 0001, South Africa}
\affiliation{Laser Research Institute, University of Stellenbosch, Stellenbosch 7602, South Africa}
\author{Thomas Konrad} \affiliation{University of Kwazulu-Natal, Private Bag X54001, 4000 Durban, South Africa}
\author{Andrew Forbes} \affiliation{CSIR National Laser Centre, PO Box 395, Pretoria 0001, South Africa}
\affiliation{University of Kwazulu-Natal, Private Bag X54001, 4000 Durban, South Africa}
\affiliation{Laser Research Institute, University of Stellenbosch, Stellenbosch 7602, South Africa}

\begin{abstract}
The turbulence induced decay of orbital angular momentum (OAM) entanglement between two photons is investigated numerically and experimentally. To compare our results with previous work, we simulate the turbulent atmosphere with a single phase screen based on the Kolmogorov theory of turbulence. We consider two different scenarios: in the first only one of the two photons propagates through turbulence, and in the second both photons propagate through uncorrelated turbulence. Comparing the entanglement evolution for different OAM values, we found the entanglement to be more robust in turbulence for higher  OAM values. We derive an empirical formula for the distance scale at which entanglement decays in term of the scale parameters and the OAM value.
\end{abstract}


\maketitle

\section{Introduction} %

Laguerre-Gaussian (LG) modes are currently the focus of intense research within the quantum information community. While the polarization of light is a two-level system for single photons, the LG modes constitute a multi-level system, which provides the possibility to store and process photonic {\it qudits} \cite{MT}, i.e.\ a superposition of a multitude of independent (orthogonal) states in a single photon. An LG beam with azimuthal mode index $\ell$ carries an orbital angular momentum (OAM) of $\ell\hbar$ per photon \cite{ABSW,FHRD}. One can, in principle, use the OAM modes of light to implement a higher dimensional state space for a single photon \cite{MMcLaren2012, ACDada2011}. A pair of photons entangled in their OAM degree of freedom can be generated by spontaneous parametric down conversion (SPDC).

In view of the increased storage capacity per photon and the corresponding new potential quantum communication protocols, OAM entangled photons are candidate carriers for long-range quantum communication \cite{P2} and higher dimensional quantum key distribution \cite{SG}. So far, protocols to achieve long-range quantum communication are based on the distribution of quantum correlations (entanglement) between the nodes of a network combined with teleportation of these correlations from one node to the next, a process also known as entanglement swapping \cite{eswap, eswap2, eswap3}. Whether long-range quantum communication can be successfully established depends on the ability to distribute entanglement efficiently over medium distances. 

One of the biggest challenges that confronts the use of OAM photon states for quantum communication is the distortion of the modes during transmission over large distances. OAM encoding is incompatible with single mode optical fiber, because it only supports modes with zero OAM. An alternative is to use free-space communication. However, OAM modes suffer distortion due to the scintillation process that the photon pair experiences while propagating through the turbulent atmosphere, which negatively affects their entanglement. Moreover, in a practical communication system the information would be encoded in terms of a finite number of OAM basis elements (say for example $\{ {|{-1},{-1}\rangle}; {|{-1},1\rangle}; {|1,{-1}\rangle}; {|1,1\rangle} \}$ in the qubit case). As such the quantum information is restricted to a proper (finite dimensional) subspace of the complete OAM Hilbert space. The orthogonal compliment of this subspace does not represent any information. Although the quantum state of the photon pair is initially prepared to lie completely within the information encoding subspace, scintillation generally causes the state of the photon pair to be partially transferred to the orthogonal compliment. The result is a loss of information in the photon field. (Note that this is different from the usual decoherence process where information is transferred to the environment, which is assumed to form a tensor product with the information-carrying quantum system. In the OAM case the information is lost to a part of the same Hilbert space. This is a drawback of OAM based systems compared to polarization based systems, because in the latter case the entire Hilbert space is used to encode the information).

Previous theoretical studies of the effects of atmospheric turbulence on the OAM modes have considered the effect of turbulence on the detection probability of OAM modes \cite{P1b, P2, TB}, the attenuation and crosstalk among multiple OAM channels \cite{JA} and the decay of entanglement for bipartite qubits \cite{P1,FSR}. These studies are all based on the Paterson model using a single phase screen \cite{P2} with the exception of the analytical study presented in \cite{FSR} and the numerical study in \cite{JA} which are both based on a multiple phase screen approach. 

The random phase function in the single phase screen model represents the turbulence according to the Kolmogorov theory \cite{K,AP} as parametrized by the Fried parameter \cite{FRI}
\be
r_0=0.185 \left( \frac{\lambda^2}{C_n^2 z} \right)^{3/5},
\label{eqn:fried}
\ee
where $C_n^2$ is the refractive index structure constant, $z$ is the propagation distance and $\lambda$ is the wavelength. This model has also been used to simulate turbulence in experimental studies. For instance, the crosstalk among OAM modes was experimentally measured \cite{MehulMalik2012, BRodenburg2012a}, where the turbulence was simulated with a single phase screen. 

Other experimental studies not based on the single phase screen approximation include the work by Pors {\it et al.} \cite{BJP}, where it was shown, using coincidence counts, that the number of entangled modes (the Shannon dimensionality) decreases with increasing scintillation and the recent work by Rodenburg {\it et al.} \cite{Rodenburg2013} where a thick turbulent medium was simulated in the lab with two phase screens and the cross-talk in the communication channel is reduced using an adaptive correction of the turbulence as well as optimization of the channel encoding.

To date, the only experimental study directly addressing the dissipation of OAM entanglement due to atmospheric turbulence was reported in  \cite{Ibrahim2012}. Our aim here is to expand on that study. We investigate numerically and experimentally the decay of OAM entanglement of photon pairs propagating in a turbulent atmosphere modelled with a single phase screen. We particularly focus on the effect of turbulence on different OAM modes. We use the Kolmogorov theory of turbulence \cite{K,AP} and restrict our analyses to the two-level (qubit) case. The quantum entanglement is quantified in terms of Wootter's concurrence \cite{W}. We compare our results with two theories predicting the evolution of OAM entanglement in atmospheric turbulence: the results presented by Smith and Raymer (S\&R) in \cite{P1} and the infinitesimal propagation equation (IPE) derived in \cite{FSR}.

This paper is organized as follows: we give a theoretical background in Sec.~\ref{sec:thback} and then describe the numerical procedures in Sec.~\ref{sec:numsim} followed by a description of the experimental procedure in Sec.~\ref{sec:exp}. Our results are presented and discussed in Sec.~\ref{sec:results}. Some conclusions are provided in Sec.~\ref{sec:con}.

\section{Theoretical background}
\label{sec:thback}

The effect of atmospheric turbulence on a classical optical beam has been the subject of many books \cite{AP, AW, Tat, Lukin2002}. Here we briefly outline the salient points.

It is reasonable to assume that the field representing the photons propagating through the turbulent atmosphere is paraxial and uniformly polarized. Moreover, the refractive index fluctuation in a turbulent atmosphere is much smaller than the average refractive index (which is approximately equal to 1). Under these conditions the propagation of the photon field is given by a linear paraxial wave equation with an additional noise term, which contains the refractive index fluctuation \cite{AP}
\begin{equation}
\nabla_T^2 g({\bf x}) - i 2 k_0 \partial_z g({\bf x}) + 2 \delta n({\bf x}) k_0^2 g({\bf x}) = 0 ,
\label{pgv}
\end{equation}
where $\nabla_T^2 = \partial^2/\partial x^2 + \partial^2/\partial y^2$, $k_0$ is the wave number in vacuum, ${\bf x} = x\hat{x} + y \hat{y} + z \hat{z}$ is a three-dimensional position vector and the refractive index fluctuation is given by $\delta n ({\bf x})=n-1$. The scalar field $g({\bf x})$ is related to the electric field by 
\begin{equation}
{\bf E}({\bf x}) = \hat{n} g({\bf x}) \exp(-i k_0 z) ,
\end{equation}
where $\hat{n}$ represents a uniform polarization vector. Propagation is assumed to be along the $z$-direction.

One can see from Eq.~(\ref{pgv}) that the effect of the turbulent atmosphere on the scalar field is completely determined by the properties of the refractive index fluctuation $\delta n ({\bf x})$.
The effect of the turbulence comes in the form of random phase modulations that are continuously introduced along the propagation path. In the single phase screen approximation this continuous modulation process is replaced by a single phase distortion. That is, we assume that the whole atmospheric medium can be replace with a single phase screen.

\begin{figure}[th]
\begin{center}
\includegraphics{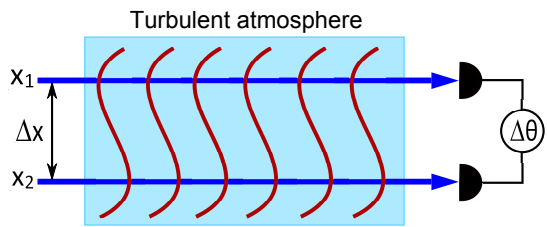}
\caption{(Color
  online) A method for measuring the phase differences between two coherent beams propagating in a turbulent atmosphere. The phase difference is measured with an interferometer.}
\label{fig:interference}
\end{center}
\end{figure}

The random phase is related to the refractive index fluctuation through
\begin{equation}
\theta({\bf X}) = k_0 \int_{0}^{\Delta z} \delta n({\bf x})\ {\rm d} z ,
\label{fasefunk}
\end{equation}
where $\Delta z$ represents the propagation distance through the turbulence and ${\bf X} = x\hat{x} + y \hat{y}$ is the two-dimensional position vector. To find the expression for the random phase function one can envisage an experiment to measure the phase difference between the output optical fields obtained after two parallel coherent optical beams are sent through the turbulence separated by a certain distance $\Delta x$ as illustrated in Fig.~\ref{fig:interference}. The interference between these two beams, which depends on the difference in phase $\Delta \theta$, can then be used to calculate the phase structure function given by 
\begin{eqnarray} \nonumber
D_{\theta}({\bf X}_1-{\bf X}_2) &=& \left\langle[ \left(\theta({\bf X}_1) -  \theta({\bf X}_2 \right) ]^2 \right\rangle \\ 
&=& 2 \left[B_{\theta}(0) - B_{\theta}({\bf X}_1-{\bf X}_2) \right].
\label{eqn:phstruct}
\end{eqnarray}
The last expression in Eq.~(\ref{eqn:phstruct}) relates the phase structure function to the phase autocorrelation function given by 
\begin{equation}
B_{\theta}({\bf X}_1-{\bf X}_2) = \left\langle \theta({\bf X}_1) \theta({\bf X}_2) \right\rangle .
\end{equation}
Note that due to the homogeneous statistical properties of the phase functions the phase autocorrelation function only depends on the relative coordinates. In fact, since the phase functions are also isotropic the phase autocorrelation function actually only depends on the magnitude of the relative coordinates.
The definition of the phase in Eq.~(\ref{fasefunk}) ignores an overall constant phase related to the average refractive index, which cancels in the interference and therefore does not contribute to the correlation function. So the phase autocorrelation function becomes 
\begin{equation}
B_{\theta}({\bf X}_1-{\bf X}_2) = k_0^2 \iint_{0}^{\Delta z} \left\langle \delta n({\bf x}_1) \delta n({\bf x}_2) \right\rangle {\rm d} z_1 {\rm d} z_2 ,
\label{bkor}
\end{equation}
which gives a relationship between the phase autocorrelation function and the refractive index autocorrelation function. The integrant in Eq.~(\ref{bkor}) is the refractive index autocorrelation function $B_n$ which is related to the refractive index structure function $D_n$ through an expression analogous to Eq.~(\ref{eqn:phstruct}). Thus 
\begin{equation}
B_n(r) = \left\langle \delta n({\bf x}_1) \delta n({\bf x}_2) \right\rangle = B_n(0) - \frac{1}{2} D_n(r),
\label{bfunk}
\end{equation}

where the refractive index structure function is given by \cite{AP}
\begin{equation}
D_n(r) = C_n^2 r^{2/3} = \left\langle \left[ \delta n({\bf x}_1) - \delta n({\bf x}_2)\right]^2 \right\rangle
\label{dfunk}
\end{equation}
with $C_n^2$ being the refractive index structure constant of Eq.~(\ref{eqn:fried}) and $r=|{\bf x}_1-{\bf x}_2|$.

From the Wiener-Khinchin theorem it now follows that the refractive index power spectral density is given by the three-dimensional Fourier transform of the refractive index autocorrelation function
\begin{equation}
\Phi_n({\bf k}) = \iiint^{\infty}_{-\infty} \left\langle \delta n(0) \delta n({\bf x}) \right\rangle \exp ({\rm i} {\bf k} \cdot {\bf x})\ {\rm d}^3 x.
\label{wienerk}
\end{equation}
Using Eqs.~(\ref{dfunk}) and (\ref{bfunk}) in Eq.~(\ref{wienerk}) one obtains the Kolmogorov power spectral density for the refractive index fluctuation given by  \cite{AP}
\begin{equation}
\Phi_n(k) = 0.033 C_n^2 k^{-11/3} ,
\label{kolspek}
\end{equation}
where $k$ is the magnitude of the three dimensional coordinate vector in the Fourier domain.

One can use the expression in Eq.~(\ref{wienerk}) to infer an expression for the random function of the refractive index fluctuation. Such a random function is conveniently defined in terms of its inverse Fourier transform
\begin{equation}
\delta n({\bf x}) = \iiint^{\infty}_{-\infty} \tilde{\chi}({\bf k}) \left[ \frac{\Phi_n({\bf k})}{\Delta_k^3} \right]^{1/2} \exp (-{\rm i} {\bf k} \cdot {\bf x})\ \frac{{\rm d}^3 k}{(2\pi)^3}
\label{dnfunk}
\end{equation}
where $\tilde{\chi}({\bf k})$ is a normally distributed random complex spectral function and $\Delta_k$ is its correlation width in frequency domain. The latter is inversely proportional to the spatial extent of the refractive index fluctuation (typically given by the outer scale of the turbulence). Since the refractive index fluctuation $\delta n$ is an asymmetric real-valued function $\tilde{\chi}^*({\bf k})=\tilde{\chi}(-{\bf k})$. The autocorrelation function of the random function is given by
\begin{equation}
\langle \tilde{\chi}({\bf k}_1) \tilde{\chi}^*({\bf k}_2) \rangle = (2\pi\Delta_k)^3 \delta_3 ({\bf k}_1-{\bf k}_2) .
\label{verwrand}
\end{equation}
One can readily verify that Eq.~(\ref{dnfunk}) is consistent with Eq.~(\ref{wienerk}).

Next we substitute Eq.~(\ref{dnfunk}) into Eq.~(\ref{bkor}). Using Eq.~(\ref{verwrand}) to evaluate the ensemble average and evaluating one of the three dimensional Fourier integrals we arrive at
\begin{eqnarray}
B_{\theta}({\bf X}_1-{\bf X}_2) & = & k_0^2 \iiint^{\infty}_{-\infty} \Phi_n({\bf k}_1)\iint_{0}^{\Delta z} \exp (-{\rm i} {\bf k}_1 \cdot {\bf x}_1) \nonumber \\
& & \times  \exp ({\rm i} {\bf k}_1 \cdot {\bf x}_2)\ {\rm d} z_1\ {\rm d} z_2\ {{\rm d}^3 k_1\over (2\pi)^3} ,
\label{ff0}
\end{eqnarray}
where we used the symmetry of the power spectral density $\Phi_n(-{\bf k})=\Phi_n({\bf k})$.

Evaluating the two $z$-integrals we obtain
\begin{equation}
\iint_{0}^{\Delta z} \exp \left[-{\rm i} k_z (z_1-z_2) \right]\ {\rm d} z_1\ {\rm d} z_2 = {2\over k_z^2} [1-\cos(k_z \Delta z)] .
\label{zint}
\end{equation}
Since $\delta n \ll 1$, the effect of the turbulent atmosphere on light propagating through it requires a long propagation distance to become significant. This propagation distance is much longer than the correlation distance of the turbulent medium. Therefore one can assume that $\Delta z$ is much larger than this correlation distance, which implies that the result in Eq.~(\ref{zint}) acts like a Dirac delta function. One can therefore substitute $k_z=0$ in $\Phi_n$ in Eq.~(\ref{ff0}) and pull $\Phi_n$ out of the $k_z$-integral. The integral over $k_z$ then gives
\begin{equation}
\int^{\infty}_{-\infty} {2\over k_z^2} [1-\cos(k_z \Delta z)]\ {\rm d} k_z = 2 \pi \Delta z .
\label{zint0}
\end{equation}
The resulting expression for the phase autocorrelation function is then \cite{MF1, KNEPP1983}
\begin{eqnarray}
B_{\theta}({\bf X}_1-{\bf X}_2) & = & \left\langle \theta({\bf X}_1) \theta({\bf X}_2) \right\rangle \nonumber \\
& = & {k_0^2 \Delta z} \iint^{\infty}_{-\infty} \exp [ -{\rm i} {\bf K} \cdot ({\bf X}_1-{\bf X}_2)] \nonumber \\
& & \times \Phi_n({\bf K},0)\ {{\rm d}^2 K\over (2\pi)^2} .
\label{ff1}
\end{eqnarray}

We now use the expression in Eq.~(\ref{ff1}) to infer an expression for the random phase function, similar to the way we obtained the expression for $\delta n$ in Eq.~(\ref{dnfunk}). The expression is
\begin{eqnarray}
\theta({\bf X}) & = & {k_0\over\Delta_k} \iint^{\infty}_{-\infty} \tilde{\xi}({\bf K}) \left[ {\Delta z \Phi_n({\bf K},0)}\right]^{1/2} \nonumber \\
& & \times \exp (-{\rm i} {\bf K} \cdot {\bf X})\ {{\rm d}^2 K\over (2\pi)^2}
\label{phfunk}
\end{eqnarray}
where $\tilde{\xi}({\bf K})$ is a two-dimensional normally distributed random complex spectral function such that
\begin{equation}
\langle \tilde{\xi}({\bf K}_1) \tilde{\xi}^*({\bf K}_2) \rangle = (2\pi\Delta_k)^2 \delta_2 ({\bf K}_1-{\bf K}_2) .
\label{verwrand2}
\end{equation}  
One can now verify that Eq.~(\ref{phfunk}) is consistent with Eq.~(\ref{ff1}).

For a real-valued, asymmetric phase function $\tilde{\xi}^*({\bf K})=\tilde{\xi}(-{\bf K})$, however, in the numerical simulation one normally uses a completely asymmetric two-dimensional random complex function $\tilde{\xi}({\bf K})$, which implies that the resulting phase function is complex  \cite{MF1, KNEPP1983}
\begin{eqnarray}
\theta_1({\bf X}) + {\rm i} \theta_2({\bf X}) & = & {k_0\over\Delta_k} \Delta z^{1/2} \nonumber \\
& & \times {\cal F}^{-1} \left\{ \tilde{\xi}({\bf K}) \Phi_n({\bf K},0)^{1/2} \right \} ,
\label{phfunkc}
\end{eqnarray}
where ${\cal F}^{-1}$ represents a two-dimensional inverse Fourier transform. As a result, one calculation gives two random phase functions for two phase screens, having transmission functions $t_1=\exp({\rm i} \theta_1)$ and $t_2=\exp({\rm i} \theta_2)$, respectively.

\section{Experimental procedure}
\label{sec:exp}
\begin{figure}[th]
\begin{center}
\includegraphics{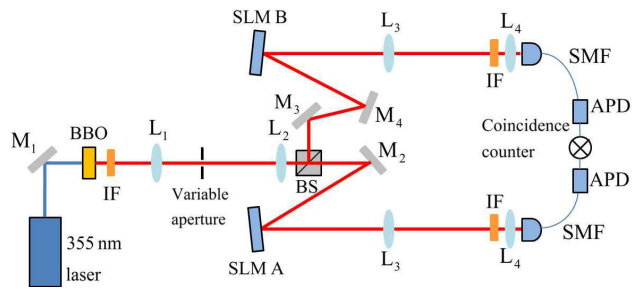}
\caption{(Color
  online) Experimental setup used to detect the OAM eigenstate after SPDC. The plane of the crystal was relayed imaged onto two separate SLMs using lenses, $\textrm{L}_{1}$ and $\textrm{L}_{2}$ ($\textrm{f}_{1} = 200$ mm and $\textrm{f}_{2} = 400$ mm), where the LG modes were selected. Lenses $\textrm{L}_{3}$ and $\textrm{L}_{4}$ ($ \textrm{f}_{3} = 500$ mm and $\textrm{f}_{4} = 2$ mm) were used to relay image the SLM planes through 10 nm bandwidth interference filters (IF) to the inputs of the single-mode fibres (SMF). The fibres were connected to avalanche photodiodes (APDs), which were then connected to a coincidence counter.}
\label{fig:setup}
\end{center}
\end{figure}

Figure~(\ref{fig:setup}) shows our experimental setup in which  a 3~mm thick type I BBO crystal was pumped with a mode-locked laser source (Gaussian mode) that has a wavelength of 355~nm and an average power of 350~mW to produce collinear, degenerate entangled photon pairs via SPDC. The crystal plane was imaged using a $4f$ telescope with $\textrm{L}_{1}$ ($f_{1} = 200$~mm) and $\textrm{L}_{2}$ ($f_{2} = 400$~mm) onto two separate spatial light modulators (SLMs). The LG modes to be measured, together with the turbulence were encoded onto the SLMs. A second $4f$ telescope with $\textrm{L}_{3}$ ($f_{3} = 500$~mm) and $\textrm{L}_{4}$ ($f_{4} = 2$~mm)  was used to re-image the SLM planes to the inputs of the single-mode fibers, where only the fundamental Gaussian modes were coupled into the fibers. The fibers were connected to avalanche photodiodes (APDs), which were then connected to a coincidence counter where the photon pairs were registered. The fluctuations from the pump beam produced an uncertainty in the measured coincidence counts of approximately $5\%$. All measured coincidence counts were accumulated over a 10 second integration time with a gating time of 12~ns.

The atmospheric turbulence was simulated by adding random phase fluctuations [as given by Eq.~(\ref{phfunkc})] to the phase function of one of the SLMs in the case when only one of the photons was propagated through turbulence, and to the phase functions of both SLMs in the case when both photons were propagated through turbulence. The scintillation strengths (${\rm w}_0/r_0$) ranged from 0 to 4 with an increment of 0.2. Measurements for each scintillation strength were repeated 30 times and a full state tomography was done after each run to reconstruct the density matrix. The negative eigenvalues that occur because of experimental imperfection were removed with the method described in \cite{Ibrahim2012}.  The reconstructed density matrices were then averaged to obtain the mean density matrix for each scintillation strength.

\section{Numerical simulation procedure}
\label{sec:numsim}

Without any loss of generality we consider the case where the initial state is a Bell state
\begin{equation}
|\Psi\rangle = \frac{1}{\sqrt{2}}\left(|q\rangle_A |\bar{q}\rangle_B + |\bar{q}\rangle_A |q\rangle_B \right).
\label{bell}
\end{equation}
Here $|q\rangle$ represents a single photon LG mode and $q$ is the magnitude of the azimuthal index (i.e.\ $q=|\ell|$), with $\bar{q} \equiv -q$ and the radial index is zero ($p=0$). The subscripts $A$ and $B$ are used to label the two different paths of the two photons through turbulence.

The OAM basis is represented by the LG modes $M^{\rm LG}_{\ell p}(r,\phi,t)=\langle {\bf x} | \ell, p\rangle$, which can be expressed in normalized cylindrical coordinates by
\begin{eqnarray}
M^{\rm LG}_{\ell p}(r,\phi,t) &=& {\cal N} {r^{|\ell|} \exp({\rm i} \ell \phi) (1 + {\rm i} t)^p \over (1 - {\rm i} t)^{p + |\ell| +1}} \nonumber \\   
& & \times L_p^{|\ell|}\left(\frac{2 r^2}{1 + t^2} \right) \exp \left(\frac{-r^2}{1 - {\rm i} t} \right) ,
\label{eqn:lg}
\end{eqnarray}
where $L_p^{|\ell|}$ represents the generalized Laguerre polynomials with the parameters $\ell$ and $p$ being the azimuthal and the radial mode indices, respectively; $r=(x^2+y^2)^{1/2}/{\rm w}_0$, $\phi$ is the azimuthal angle and $t=z/z_R$, with ${\rm w}_0$ being the beam waist radius, $z_R$ being the Rayleigh range ($=\pi{\rm w}_0^2/\lambda$) and $\lambda$ being the wavelength. The normalization constant is given by
\begin{equation}
{\cal N} = \left[ \frac{p! 2^{|\ell| + 1}}{\pi (p + |\ell|)!} \right]^{1/2}.
\end{equation}

\begin{figure}[ht]
\centerline{\includegraphics{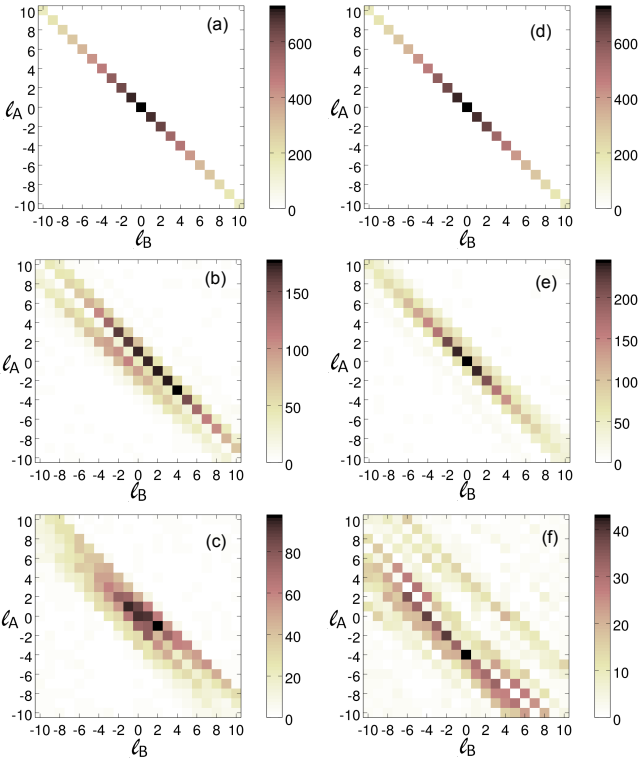}}
\caption{(Color
  online) Mode scattering under the effect of turbulence given by the coincidence counts for simultaneous measurements of modes with azimuthal index $\ell_A$ in the signal beam and $\ell_B$ in the idler beam when only one of the two photon propagates through turbulence [(a), (b) and (c)] and when both photons propagate through turbulence [(d),(e) and (f)]. With no turbulence [(a) and (d)], only anti-correlated coincidences are observed. As the scintillation strength increases to ${\rm w}_0/r_0 = 2$ [(b) and (e)] and ${\rm w}_0/r_0 = 4$ [(c) and (f)], the mode scattering becomes more pronounced.}
\label{fig:MS}
\end{figure}

When a photon with a given OAM mode propagates through turbulence, the distortions cause the photon to become a superposition of many OAM modes. In other words, any particular OAM state of the photon is scattered into many OAM states. This scattering has been measured experimentally by displaying the appropriate phase profiles on the SLMs  (Fig.~\ref{fig:setup}) and is illustrated in Fig.~\ref{fig:MS} for both scenarios. Initially [Fig.~\ref{fig:MS}(a) and (d)] there is no turbulence and we have coincidences only along the diagonal (when $\ell_A = -\ell_B$). When we turn on the turbulence, we detect more coincidences when $\ell_A \neq -\ell_B$ as we increase the scintillation strength.

Here we only consider qubits. Therefore, when we compute any density matrix, we extract only the information contained in the modes where $\ell = \pm q$ and $p = 0$. Hence, we exclude all other modes in the expression of the density matrix. 

The state of photon A or B changes as follows after propagating over a distance of $\Delta z$ through turbulence
\begin{eqnarray}
|q\rangle_A       & \rightarrow & a_q |q\rangle_A + a_{\bar{q}} |\bar{q}\rangle_A \nonumber \\
|\bar{q}\rangle_A & \rightarrow & b_q |q\rangle_A + b_{\bar{q}} |\bar{q}\rangle_A \nonumber \\
|q\rangle_B       & \rightarrow & c_q |q\rangle_B + c_{\bar{q}} |\bar{q}\rangle_B \nonumber \\
|\bar{q}\rangle_B & \rightarrow & d_q |q\rangle_B + d_{\bar{q}} |\bar{q}\rangle_B ,
\end{eqnarray}
where $a_q$, $a_{\bar{q}}$, etc.\ are the complex coefficients in the expansion of the distorted state in terms of the OAM basis. In other words, $a_q = \langle q|_A U_{\Delta z} |q\rangle_A$, $a_{\bar{q}} = \langle \bar{q}|_A U_{\Delta z} |q\rangle_A$, and so forth where the unitary operator $U_{\Delta z}$ represents propagation through turbulence over a distance of $\Delta z$. That is, one can express the distorted state after propagation by $|\Psi'\rangle = U_{\Delta z}|\Psi\rangle$.

After propagating through turbulence, the initial state in Eq.~(\ref{bell}) will be transformed into
\begin{eqnarray}
|\Psi\rangle \rightarrow |\Psi\rangle^{out} & = & C_1 |q\rangle_A |q\rangle_B + C_2 |q\rangle_A |\bar{q}\rangle_B \nonumber \\
& & + C_3 |\bar{q}\rangle_A |q\rangle_B + C_4 |\bar{q}\rangle_A |\bar{q}\rangle_B ,
\label{projtra}
\end{eqnarray}
where
\begin{eqnarray} \nonumber
C_1 &=& \frac{1}{\sqrt{2}} c_q, \nonumber \\
C_2 &=& \frac{1}{\sqrt{2}} a_q,  \nonumber \\
C_3 &=& \frac{1}{\sqrt{2}}c_{\bar{q}}, \\
C_4 &=& \frac{1}{\sqrt{2}}a_{\bar{q}},\nonumber
\end{eqnarray}
in the case where only photon A is propagated through turbulence, and

\begin{eqnarray} \nonumber
C_1 &=& \frac{1}{\sqrt{2}}\left(a_q d_q + b_q c_q \right), \nonumber \\
C_2 &=& \frac{1}{\sqrt{2}}\left(a_q d_{\bar{q}} + b_q c_{\bar{q}} \right), \nonumber \\
C_3 &=& \frac{1}{\sqrt{2}}\left(a_{\bar{q}} d_q + b_{\bar{q}} c_q \right), \\
C_4 &=& \frac{1}{\sqrt{2}}\left(a_{\bar{q}} d_{\bar{q}} + b_{\bar{q}} c_{\bar{q}}\right), \nonumber
\end{eqnarray}
in the case where both photons are propagated through turbulence.

Note that, since only a restricted set of basis elements are retained, the transformation in Eq.~(\ref{projtra}) is not unitary ($|\Psi\rangle^{out}\neq U_{\Delta z}|\Psi\rangle$). The transformed state after the propagation $|\Psi\rangle^{out}$ is however still a pure state, but it is obtained for a specific instance of the turbulent medium (or, in the case of the numerical simulation, for specific phase functions on the phase screens). We do not assume that we have detailed knowledge of the medium. Therefore, one needs to compute the ensemble average of the density matrix over all possible (or a representative set of) instances of the medium (or of the phase functions). The resulting density matrix becomes that of a mixed state. This mixture can be seen as the result of `tracing over the environment.' The mean density matrix is then given by 

\begin{equation}
\rho =  \frac{\sum^N_n |\Psi_n\rangle \langle \Psi_n|}{{\rm Tr} \left\{ \sum^N_n |\Psi_n\rangle \langle \Psi_n| \right\} },
\label{normdens}
\end{equation}
where $|\Psi_n\rangle$ is the state of the qubit after the photons propagate through the $n^{th}$ phase screen (the $n^{th}$ realization of the turbulence medium).

The concurrence, which is used as a measure of entanglement \cite{W}, is given by
\begin{equation}
C(\rho) = \max\{0,\sqrt{\lambda_1} - \sqrt{\lambda_2} - \sqrt{\lambda_3} - \sqrt{\lambda_4}\} ,
\end{equation}
where $\lambda_i$ are the eigenvalues, in decreasing order, of the Hermitian matrix
\begin{equation}
R = \rho(\sigma_y\otimes\sigma_y)\rho^*(\sigma_y\otimes\sigma_y) ,
\label{eqn:R}
\end{equation}
with $^*$ representing the complex conjugate and $\sigma_y$ being the Pauli $y$-matrix
\begin{equation}
\sigma_y =  \left[ \begin{array}{rr} 0 & {\rm -i} \\ {\rm i} & 0 \end{array} \right].
\end{equation}

To simulate the propagation of an entangled quantum state one needs to propagate each of the separate components that make up the state. For the Bell state in Eq.~(\ref{bell}), this implies two optical fields for each of the propagation paths. Hence, four propagation simulations for each run. The four input optical fields are produced as $256\times 256$ arrays of samples of the complex function that represents the mode in the input plane of the system. The complex function for the modes are given in Eq.~(\ref{eqn:lg}), where we set $\ell=\pm q$, $p=0$ and $z=0$. We consider the different cases where $q=1,3,5$ and 7. In the simulation, we first multiply the optical fields with the transmission function representing the random phase computed in Eq.~\ref{phfunkc}. Then the resulting fields are propagated through free-space over a distance of $\Delta z$.

After each free-space propagation step the density matrix of the resulting quantum states is determined by extracting the coefficients of the different modes from the four fields at that point and combining these coefficients into the expression for the states according to Eq.~(\ref{projtra}). 

One such run gives a sequence of pure states that represents the evolution of the quantum state of the pair of photons as it propagated through a specific simulated turbulent atmosphere. We performed a number ($N=1000$) of such runs for $N$ different simulated turbulent atmospheres to obtain $N$ different evolutions of the quantum state. These $N$ runs are used to perform ensemble averaging for each of the elements in the evolution sequence, as expressed in Eq.~(\ref{normdens}), to obtain a sequence of density matrices that represent the evolution of the bi-photon state from an initial pure state to the mixed quantum state that one would observe at a particular point along the propagation path.

\section{Results and discussion}
\label{sec:results}

\subsection{Theoretical models}
\label{sec:pts}

We compare our results with two theories predicting the evolution of OAM entanglement in atmospheric turbulence: the results presented by Smith and Raymer (S\&R) in \cite{P1} and the infinitesimal propagation equation (IPE) derived in \cite{FSR}. The S\&R theory is based on a single phase screen approximation. That is, it assumes that the overall effect of the turbulent medium can be represented by a single phase distortion on the beam. This approximation limits the validity of the S\&R prediction to the weak fluctuations regime. Furthermore, the S\&R theory makes use of the quadratic approximation of the structure function \cite{Wandzura1980} in the calculation of the phase correlation function. 

The IPE on the other hand is a first order differential equation describing the evolution of OAM entanglement in turbulence. It was derived by treating the distortion that an OAM state experiences due to propagation through a thin sheet of turbulent atmosphere as an infinitesimal transformation. It is thus based on multiple phase screens and predicts the evolution of entanglement even in the strong fluctuation regime. 

In both these studies, the concurrence is plotted against the scintillation strength represented by 
\be
\frac{{\rm w}_0}{r_0} = 5.4054{\rm w}_0 \left(\frac{C_n^2 z}{\lambda^2} \right)^{3/5},
\ee
where $r_0$ is the Fried parameter (Eq.~\ref{eqn:fried}).
This quantity depends on both the refractive-index structure constant $C_n^2$ which is a measure of the strength of the refractive-index inhomogeneities and the propagation distance $z$. It is thus a measure of the scintillation strength.

\subsection{Single photon case}
\label{sec:single}

Here we consider the case where only one of the two photons propagates through atmospheric turbulence; we refer to this as the single photon case. 

\begin{figure}[th]
\begin{center}
\includegraphics{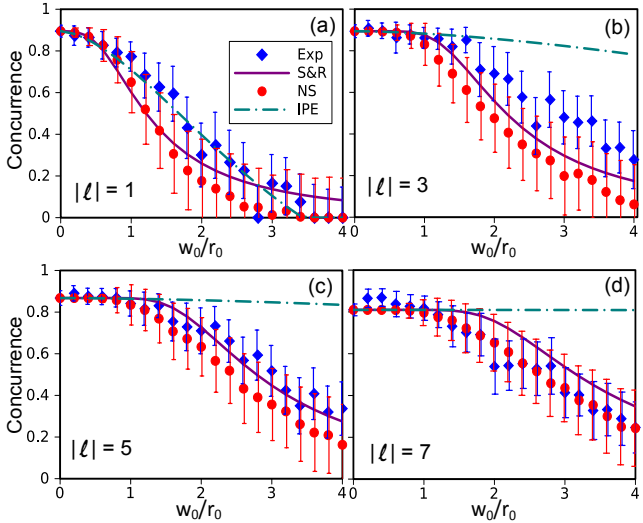}
\caption{(Color
  online) The concurrence against the scintillation strength ($w_0/r_0$) when only one photon is propagated through turbulence. In (a) $|\ell| = 1$, in (b) $|\ell| = 3$, in (c) $|\ell| = 5$ and in (d) $|\ell| = 7$. In the legend, Exp: experimental data points, 	S\&R: theory curve derived by Smith and Raymer in \cite{P1}, IPE: the infinitesimal propagation equation presented in \cite{FSR} and NS: Numerical data points.}
\label{F1}
\end{center}
\end{figure}

In Fig.~\ref{F1}, we compare the experimental data (Exp) with the numerical simulation results (NS) and the two theories discussed above, namely the S\&R theory and IPE.

Within experimental errors, the experimental results agree with the numerical results and both theories when $|\ell| = 1$. As the value of $|\ell|$ increases, the experimental results remain consistent with the numerical results and the S\&R theory but increasingly disagree with the IPE. 

We observe in Fig.~\ref{F1} that both the S\&R theory and the numerical results predict that the concurrence takes longer to decay for higher values of $|\ell|$.

The IPE also predicts that the concurrence will last longer for higher $|\ell|$-values, however, it predicts a much slower decay rate for the concurrence and it completely deviates from the other curves when $\ell>1$. The reason for this is discussed below (Sec.~\ref{ssec:IPE}).

Our experimental results also suggest that the concurrence lasts longer for higher $|\ell|$-values as we can see a clear increment between $|\ell| = 1$ and $|\ell| = 3$ as predicted by both theories and the numerical simulation. For instance when $|\ell| = 1$, the concurrence decays to zero around the point where ${\rm w}_0/r_0 = 4$ whereas the value of the concurrence is about 0.25 at ${\rm w}_0/r_0 = 4$ when $|\ell| = 3$. However there is no clear distinction between the points corresponding to $|\ell| = 3, 5$ and 7 (the concurrence is about 0.25 at ${\rm w}_0/r_0 = 4$ for all these cases). This might be due to experimental imperfection. As the $|\ell|$-value increases, it becomes more difficult to measure the mode accurately.

\subsection{Two-photon case}
In a practical quantum communication system one would need to send both photons through turbulence. For instance, we can think of a situation where a pair of entangled photons is generated and sent to two different parties (Alice and Bob) for a quantum information task such as quantum teleportation. It is thus important to study the effects of turbulence on the OAM entanglement when both photons are propagated through turbulence. We refer to this situation as the two-photon case.

\begin{figure}[th]
\begin{center}
\includegraphics{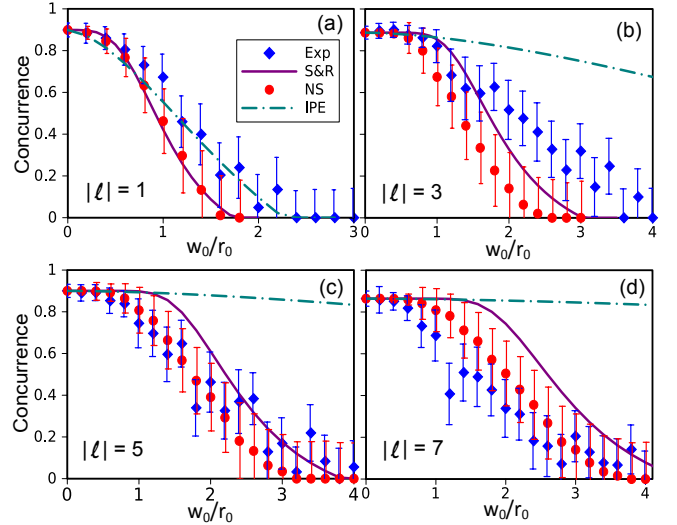}
\caption{(Color
  online) The concurrence against the scintillation strength ($w_0/r_0$) when both photons are propagated through turbulence. In (a) $|\ell| = 1$, in (b) $|\ell| = 3$, in (c) $|\ell| = 5$ and in (d) $|\ell| = 7$. In the legend, Exp: experimental data points, 	S\&R: theory curve derived by Smith and Raymer in \cite{P1}, IPE: the infinitesimal propagation equation presented in \cite{FSR} and NS: Numerical data points.}
\label{fig:two}
\end{center}
\end{figure}

Figure~\ref{fig:two} shows the evolution of the concurrence as the scintillation strength increases. The general pattern is quite similar to what was observed in the single photon case with the main difference that the concurrence decays much quicker. Here too the experimental results agree with the numerical results and both theories when $\ell = 1$ [Fig.~\ref{fig:two}(a)]. As we increase the value of $\ell$, the experimental results remain consistent with the numerical simulation and the S\&R theory but increasingly disagree with the IPE.

It can also be seen in Fig.~\ref{fig:two} that the concurrence decays slower for higher $\ell$-values, as in the single photon case. This is more clearly seen in Fig.~\ref{fig:log0} (c) and (d) where S\&R theory and the numerical simulation are plotted against the scintillation strength on a logarithmic scale.
 The IPE predicts a slower decay rate and completely deviates from the other curves when $\ell>1$ for the reasons discussed in Sec.~\ref{ssec:IPE}.

Our experimental results support the fact that the concurrence lasts longer for higher $|\ell|$-values. The concurrence decays to zero around ${\rm w}_0/r_0 = 2.5$ when $|\ell| = 1$ whereas it decays to zero around ${\rm w}_0/r_0 = 4$ when $|\ell| = 3,5$ and 7. There is no clear distinction between the points corresponding to $\ell = 3, 5$ and $\ell = 7$. This is again due to experimental imperfections.

\subsection{Scale at which entanglement decays}

\begin{figure}[th]
\begin{center}
\includegraphics{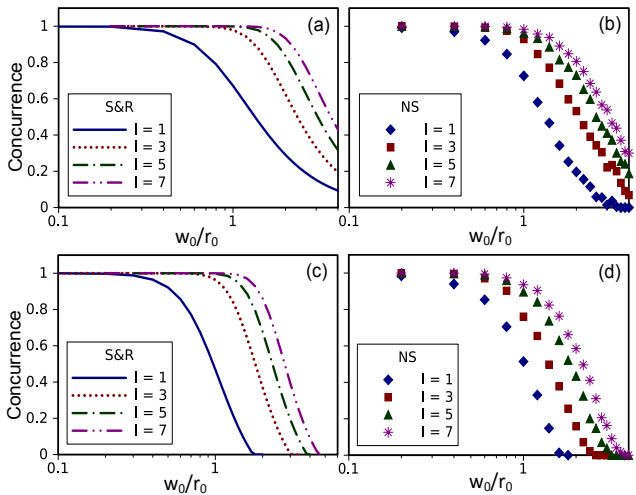}
\caption{(Color
  online) The concurrence against the scintillation strength ${\rm w}_0/r_0$ for the S\&R theory and the numerical results in the single photon case [(a) and (b)] and in the two photons case [(c) and (d)]. The horizontal axis is plotted on a logarithmic scale.}
\label{fig:log0}
\end{center}
\end{figure}

The S\&R theory predicts that the concurrence lasts longer for higher values of $|\ell|$, and that the spacing between adjacent curves decreases as $|\ell|$ increases. This is also true for the numerical simulation and can be seen in Fig.~\ref{fig:log0} where we plot the S\&R theory and the numerical results against the scintillation strength on a logarithmic scale. The fact that the concurrence survives longer for higher $|\ell|$-values suggests that the scale of entanglement decay will occur around a different point for  larger values of $\ell$:  the scale at which decoherence occurs depends on the value of $\ell$.

\begin{figure}[th]
\begin{center}
\includegraphics{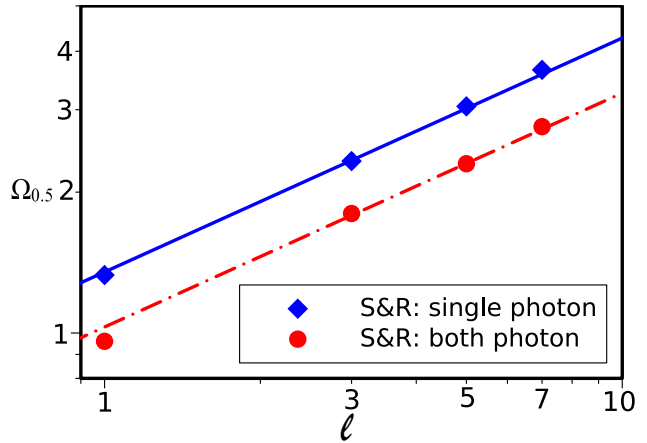}
\caption{(Color
  online) The scintillation strength against $\ell$ on a logarithmic scale for both the single photon case (diamond dots) and the two photons case (circular dots). The equation of the fitted lines are $\log_{10} \left(\Omega_{0.5} \right) = 0.5 \log_{10}(\ell) + 0.1303$ in the single photon case and $\log_{10} \left(\Omega_{0.5} \right) = 0.5 \log_{10}(\ell) + 0.01284$ in the two photon case.}
\label{fig:log}
\end{center}
\end{figure}

To find that $\ell$ dependence, we use the S\&R theory to find the values of ${\rm w}_0/r_0$ where the concurrence is equal to 0.5  for the different $|\ell|$-values considered. We'll denote them by $\Omega_{0.5}$. The result obtained is shown in Fig.~\ref{fig:log} where the values of $\Omega_{0.5}$ are plotted against the corresponding values of $\ell$ on a logarithmic scale.

We find $\Omega_{0.5} = 1.35 \sqrt{\ell}$ in the single photon case and $\Omega_{0.5} = 1.03 \sqrt{\ell}$ in the two photon case. Thus in both cases the entanglement decay  happens within an order of magnitude around the point where $\Omega_{0.5} \approx \sqrt{\ell}$. By using the expression of the Fried parameter [Eq.(\ref{eqn:fried})],
we find that the distance scale at which OAM entanglement decays as a function of $\ell$ is
\be
L_{ \rm dec} (\ell) \approx \frac{0.06 \lambda^2 \ell^{5/6}}{{\rm w}_0^{5/3} C_n^2}.
\label{eqn:Ldec}
\ee
Thus for a practical free-space quantum communication system using OAM modes as qubits, the distance between repeaters should be shorter than $L_{\rm dec}(\ell)$. For example, if one would send OAM entangled photons in a beam with ${\rm w}_0 = 10 $ cm, a wavelength of $\lambda = 1550$ nm, on a horizontal path in moderate turbulence conditions ($C_n^2 = 10^{-15}$ $m^{-2/3}$), the entanglement between the photons will decay around the distances shown in Table~\ref{tab1} for the different values of $\ell$. 

We notice in Table~\ref{tab1} that the distance scale at which entanglement decays is relatively short even in moderate turbulence. This suggests that the OAM state of light might not be suitable for long distance free-space quantum communication. One can try to increase that distance by using a smaller beam radius, but that would increase beam divergence, which in turn reduces the received power for a given receiver aperture. The entanglement decay distance can also be increased by using adaptive optics.

\begin{table}
\caption{\label{tab1} Distance scale at which entanglement decays for OAM entangled photons in a beam with ${\rm w}_0 = 10$ cm, a wavelength of $\lambda = 1550$ nm, on a horizontal path in moderate turbulence ($C_n^2 = 10^{-15}$ ${\rm m^{-2/3}}$).}
\begin{ruledtabular}
\begin{tabular}{ccccc}
$\ell$ 		   & 1   & 3   & 5   & 7   \\ 
\hline
$L_{ \rm dec} ({\rm km})$ & 6.7 & 16.7 & 25.6 & 33.7 \\
\end{tabular}
\end{ruledtabular}
\end{table}

\subsection{Truncation problem in the IPE}
\label{ssec:IPE}
While the IPE avoids the approximations that are made in the Paterson model, on which the S\&R calculation is based, it suffers from a drawback when it comes to the effect of truncations. Both the IPE and the Paterson model can in principle represent the infinite dimensional density matrix of the photonic quantum state. In practice both need to be truncated. The effect of truncations is to remove all the backward interactions from the neglected elements to those that are included in the truncated matrix. However, in the Paterson model the single phase screen represents a much thicker medium than the infinitesimal step in the IPE. As a result the Paterson model incorporates multiple scattering in the single phase screen. This causes the coupling strengths between basis element that are further apart to be stronger in the Paterson model than they are in the IPE. The IPE cannot see the multiple scattering that would take $\ell=q$ to $\ell=-q$ via the intermediate basis elements with $|\ell|<q$ if these latter basis elements are removed from the truncated matrix. As a result the coupling between $\ell=q$ and $\ell=-q$ becomes much smaller in the IPE than the equivalent coupling in the Paterson model. The IPE therefore underestimates the coupling of different basis elements that are far apart. Due to the smaller couplings in the truncated IPE, it predicts a much slower decay rate for the concurrence than is observed experimentally.

\section{Conclusions}
\label{sec:con}
We studied the evolution of OAM entanglement in atmospheric turbulence both numerically and experimentally and we considered modes with $|\ell|$-values 1,3, 5 and 7 and we compared our results with two theories: the S\&R \cite{P1} and the IPE \cite{FSR}. We considered two different scenarios: the case where only one of the two photons is propagated through turbulence and the case where both photons are propagated through turbulence. In both these scenarios, our numerical and experimental results are consistent with the S\&R theory and suggest that modes with higher $|\ell|$-values are more robust in turbulence and could thus give an advantage in a free-space quantum communication system. However, it is also observed that modes with higher $|\ell|$-values are more difficult to measure experimentally.

The numerical results agree well with the S\&R theory and experimental results and can thus be used as a tool to predict the evolution of OAM entanglement in atmospheric turbulence in other situations.

Our numerical and experimental results disagreed with the IPE when $\ell = 3,5$ and 7. The reason for this could be the fact that the IPE doesn't take into account the effects of cross-correlation between modes with different $\ell$-values.    

We derived an expression for the scale distance at which entanglement decays as a function of $\ell$. This expression can be used to find the maximum distance over which OAM entangled photons propagate before they lose their entanglement. Using typical parameters in optical communication, we found the distance scale at which OAM entanglement decays to be relatively short even in moderate turbulence. This suggests that the OAM state of light might not be suitable for long distance free-space quantum communication.

\section*{ACKNOWLEDGEMENTS}
This work was done with the financial support of an SRP Type A grant from the CSIR.


\end{document}